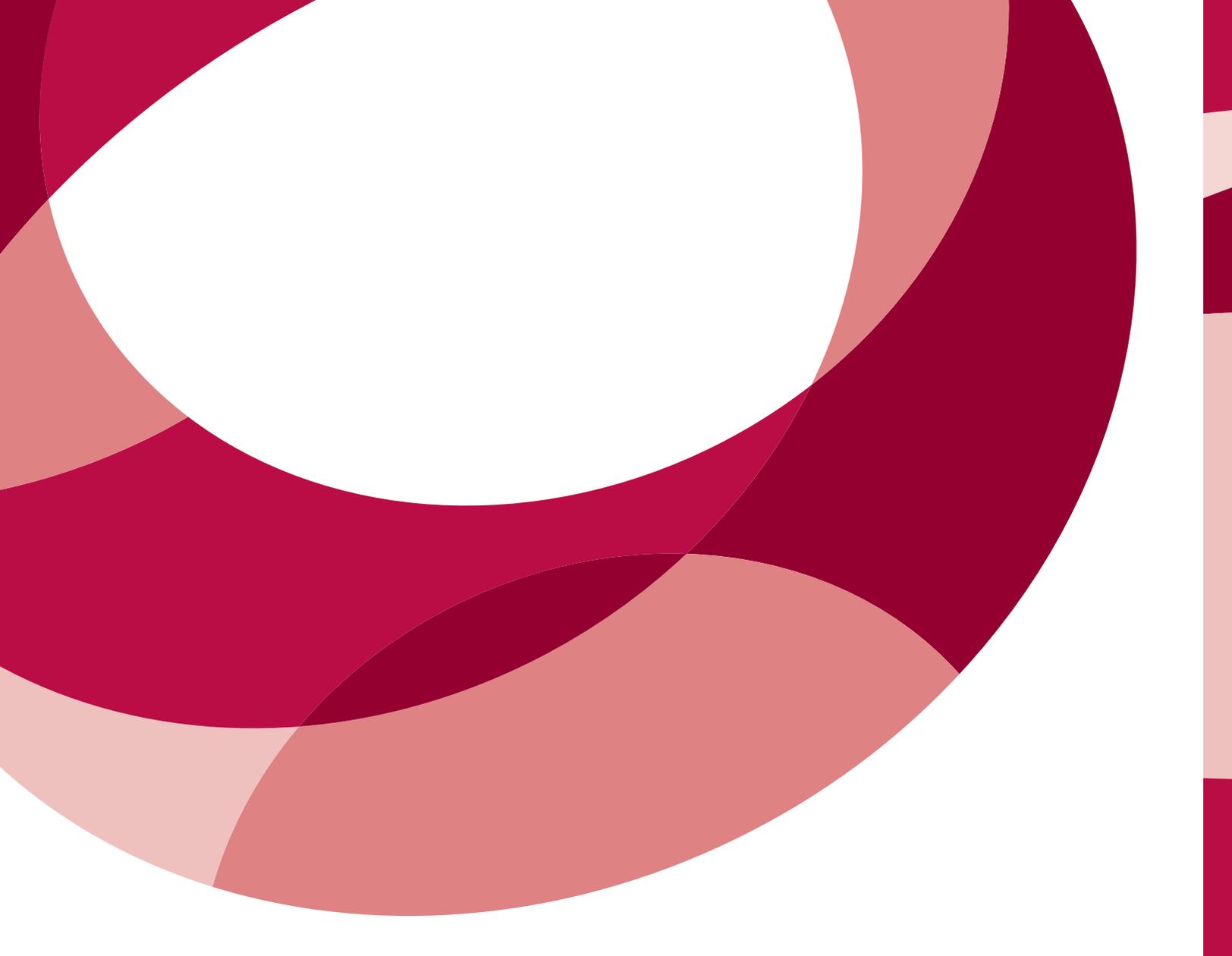

# Mathematical Foundations for Social Computing

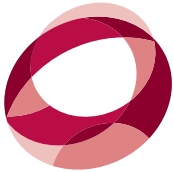

This material is based upon work supported by the
National Science Foundation under Grant No. (1136993).

Any opinions, findings, and conclusions or
recommendations expressed in this material are those of
the author(s) and do not necessarily reflect the views of
the National Science Foundation.

# Mathematical Foundations for Social Computing

Yiling Chen, Arpita Ghosh, Michael Kearns, Tim Roughgarden, and Jennifer Wortman Vaughan

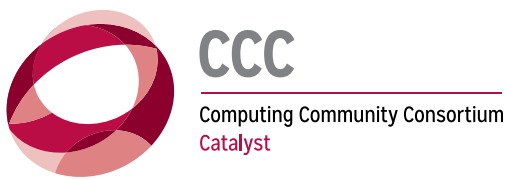
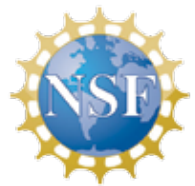

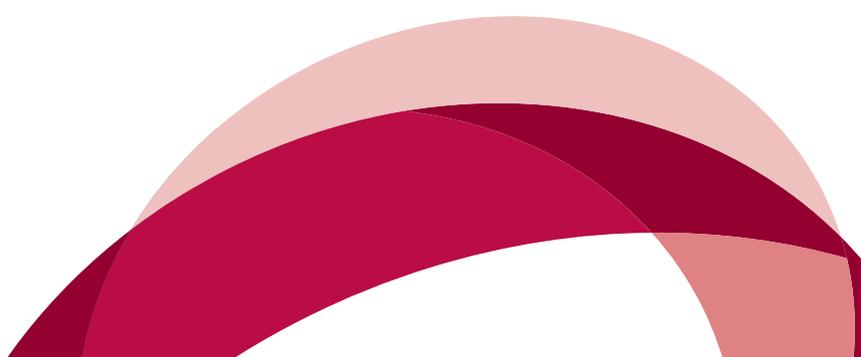

# 1 Overview

Social computing encompasses the mechanisms through which people interact with computational systems: crowdsourcing systems, ranking and recommendation systems, online prediction markets, citizen science projects, and collaboratively edited wikis, to name a few. These systems share the common feature that humans are active participants, making choices that determine the input to, and therefore the output of, the system. The output of these systems can be viewed as a joint computation between machine and human, and can be richer than what either could produce alone. The term social computing is often used as a synonym for several related areas, such as "human computation" and subsets of "collective intelligence"; we use it in its broadest sense to encompass all of these things.

Social computing is blossoming into a rich research area of its own, with contributions from diverse disciplines including computer science, economics, and other social sciences. The field spans everything from systems research directed at building scalable platforms for new social computing applications to HCI research directed towards user interface design, from studies of incentive alignment in online applications to behavioral experiments on evaluating the performance of specific systems, and from understanding online human social behavior to demonstrating new possibilities of organized social interactions. Yet a broad mathematical foundation for social computing is yet to be established, with a plethora of under-explored opportunities for mathematical research to impact social computing.

In many fields or subfields, mathematical theories have provided major contributions towards real-world applications. These contributions often come in form of mathematical models to address the closely-related problems of *analysis*—why do existing systems exhibit the outcomes they do?—and *design*—how can systems be engineered to produce better outcomes? In computer science, mathematical research led to the development of commonly used practical machine learning methods such as boosting [22] and support vector machines [11], public-key cryptography including the RSA protocol [54], widely used data structures such as splay trees [56] and techniques like locality-sensitive hashing [40], and more. Well known examples in economics include the analysis and design of matching markets [55] that have enabled Kidney Exchanges and have led to significant successes in public school admissions and residence matching for doctors and hospitals, the influence of auction theory on the design of the FCC spectrum auctions [50], and the design and redesign of the auctions used in online advertising markets [21, 61].

As in other fields, there is great potential for mathematical work to influence and shape the future of social computing. There is a small literature using mathematical models to analyze and propose design recommendations for social computing systems including crowdsourcing markets [3, 13, 26, 29, 35–37, 43, 57, 63–65], prediction markets [1, 2, 9, 10, 17, 23, 47], human computation games [41, 62], and user-generated content sites [19, 25, 28, 42]; see, for example, Ghosh [24] for a survey of one facet of this work.

However, we are far from having the systematic and principled understanding of the advantages, limitations, and potentials of social computing required to match the impact on applications that has occurred in other fields.

We note that social computing enjoys a close relationship with another emerging discipline, which is computational social science [30, 48]. But is also distinct from that field. While human and social behavior, ability, and performance are central to both, computational social science focuses primarily on the use of modern technology, data, and algorithms to understand and describe social interactions in their "natural habitats." In contrast, social computing (as the name suggests) has a much more deliberate focus on *engineering* systems that are hybrids of humans and machines, which may often entail shaping collective behavior in unfamiliar environments. Nevertheless we anticipate a continued close relationship and even blurring of the two efforts. As an example, one should expect the vast theoretical and experimental literature on the diffusion of influence and behavior in social networks to be relevant to any effort to design a social computing system which relies on such dynamics to recruit and engage workers.

In June 2015, we brought together roughly 25 experts in related fields to discuss the promise and challenges of establishing mathematical foundations for social computing. This document captures several of the key ideas discussed.

## 2 Success Stories

We begin by describing some examples in which mathematical research has led to innovations in social computing.

### 2.1 Crowdsourced Democracy

YouTube competes with Hollywood as an entertainment channel, and also supplements Hollywood by acting as a distribution mechanism. Twitter has a similar relationship to news media, and Coursera to universities. But Washington has no such counterpart; there are no online alternatives for making democratic decisions at large scale as a society. As opposed to building consensus and compromise, public discussion boards often devolve into flame wars when dealing with contentious socio-political issues. This motivates the problem of designing systems in which crowds of hundreds, perhaps millions, of individuals collaborate together to come to consensus on difficult societal issues.

Mathematical research has recently led to new systems implementing crowdsourced democracy [32]. This work builds upon a body of research in *social choice* that examines how to best take the preferences of multiple agents (human or otherwise) and obtain from them a social decision or aggregate social preference, typically accomplished through some form of voting.[1]

Consider situations where a highly structured decision must be made. Some examples are making budgets, assigning water resources, and setting tax rates. Goel et al. [32] make significant progress towards understanding the "right" mechanisms for such problems. One promising candidate is "Knapsack Voting." Recall that in the knapsack problem, a subset of items with different values and weights must be placed in a knapsack to maximize the total value without exceeding the knapsack's capacity. This captures most budgeting processes — the set of chosen budget items must fit under a spending limit, while maximizing societal value. Goel et al. [32] prove that asking users to compare projects in terms of "value for money" or asking them to choose an entire budget results in provably better properties than using the more traditional approaches of approval or rank-choice voting. Inspired by these mathematical results, Goel et al. designed a participatory budgeting platform that is fast becoming the leader for such processes in the U.S.[2] For example, this platform was recently used to decide how to spend $250,000 of infrastructure funds to improve Long Beach (CA) Council District 9, and how to allocate $2.4 million of Vallejo CA's capital improvement budget. Looking forward, it is an interesting and open research challenge to understand if these algorithms and systems yield near-optimal aggregations of societal preferences, or decisions that are near-optimal in terms of overall societal utility.

---

[1] A significant research community concerns itself primarily with *computational* social choice [6, 7]: this area has particular promise for social computing because of the problems of scale that are associated with group decision-making online, such as in crowdsourced democracy.
[2] https://pbstanford.org/cambridge/approval

## 2.2 Automated Market Makers for Prediction Markets

A *prediction market* is a market in which traders buy and sell securities with payments that are contingent on the outcome of a future event. For example, a security may yield a payment of $1 if a Democrat wins the 2016 US Presidential election and $0 otherwise. A trader who believes that the true probability of a Democrat winning the election is $p$ maximizes his expected utility by purchasing the security if it is available at a price less than $p$ and selling the security if it is available at a price greater than $p$. The *market price* of this security is thought to reflect the traders' collective belief about the likelihood of a Democrat winning.

Prediction markets have been shown to produce forecasts at least as accurate as other alternatives in a wide variety of domains, including politics [5], business [12, 58], disease surveillance [52], entertainment [51], and beyond [66], and have been widely cited by the press during recent elections. However, markets operated using traditional mechanisms like continuous double auctions (similar to the stock market) often suffer from low liquidity. Without liquidity, a market faces a chicken-and-egg problem: potential traders are dissuaded from participating due to lack of counterparties, which contributes to an even greater reduction in future trading opportunities.

To combat this problem, Hanson [34] proposed the idea of operating markets using an automated market maker called a *market scoring rule.* This market maker is an algorithmic agent that is always willing to buy or sell securities at current market prices that depend on the history of trade. Hanson's ideas build on the extensive literature on *proper scoring rules* [31], payment rules that elicit honest predictions from agents. Market scoring rules ensure that the market maker has bounded risk and that traders are unable to engage in arbitrage. Because of these desirable properties, Hanson's market scoring rules have become the prediction market implementation of choice used by companies including Consensus Point, Inkling, and Augur, and large-scale academic projects including SciCast (http://scicast.org) and the Good Judgment Project [60].

Recently there has been interest in further tapping into the informational efficiency of prediction markets and using them to obtain accurate predictions on more fine-grained events. For example, instead of viewing a Presidential election as having two possible outcomes (Democrat wins or Republican wins), we could view it as having $2^{50}$ potential outcomes, with each outcome specifying a winner in each U.S. state. Traders could then trade securities on events (combinations of outcomes) to profit on their unique knowledge, such as whether or not the same candidate will win in both Ohio and Florida, or whether or not the Republican candidate will win in at least one of Ohio, Pennsylvania, and Virginia. Such a prediction market is called a *combinatorial* prediction market. Unfortunately, due to the difficulty of keeping prices logically consistent across large outcome spaces, running market scoring rules off-the-shelf is computationally intractable for many natural examples of combinatorial markets [8].

In search of pricing rules that are tractable and preserve the logical relationships between security payoffs, Abernethy, Chen, and Vaughan [1] proposed a general framework for the design of efficient automated market makers over very large or infinite outcome spaces. They took an axiomatic approach, defining a set of formal mathematical properties that correspond to economic properties that any reasonable market should satisfy (such as "no arbitrage" and an "information incorporation" property) and fully characterized the set of pricing mechanisms that satisfy these properties. Then, using techniques from convex analysis, they provided a method for designing specific market makers that satisfy these properties. The framework enables formal reasoning of trade-offs between different economic features of these market makers as well as evaluating computational efficiency of the pricing algorithms.

This framework is particularly exciting because it offers a way to think about approximate pricing in combinatorial markets when exact pricing is still intractable. Approximate pricing for markets is challenging because approximation errors may be exploited by traders to cause the market maker to incur a large or even infinite loss. The framework of Abernethy, Chen, and Vaughan [1] characterizes deviations from exact pricing that won't add additional cost to the market maker. Building upon this understanding, Dudík et al.[15] further developed a computationally tractable method to run a large-scale prediction market that allows participants to trade almost any contract they can define over an exponentially large outcome space. This method is starting to gain traction in industry where it has been used in the PredictWise election market [16] and previous and upcoming iterations of the Microsoft Prediction Service.[3]

**2.3 Fair Division for the Masses**

Social computing systems can be used to help groups of people make decisions about their day-to-day lives. One particularly innovative example is Spliddit,[4] a website that provides tools that help groups of people achieve fair allocations. Spliddit currently offers tools to allocate rooms and divide rent payments among roommates, split taxi fares among passengers, assign credit in group projects, divide sets of (divisible or indivisible) goods among recipients, or split up tasks among collaborators. It has been featured in the New York Times[5] and had tens of thousands of users as of 2014 [33].

Spliddit's website boasts "indisputable fairness guarantees." Indeed, each of the division mechanisms employed on the site stems from the body of research on (computational) fair division [53] and comes with provable mathematical guarantees. For example, the algorithm used for room assignment and rent splitting relies on the fact that there always exists an assignment of rooms and a corresponding set of prices that is *envy-free:* every roommate prefers the room he is assigned to any other room given the prices [59]. Each roommate submits her own value for each of the rooms, under the constraint that the total value of all rooms matches the total rent for the apartment; viewed another way, each roommate is essentially submitting a proposed set of prices for each room such that she would be equally happy obtaining any room at the specified price. The algorithm then maximizes the minimum utility (value of room minus price) of any roommate subject to the constraint that envy-freeness is satisfied. The solution is also *Pareto efficient*, meaning there is no other allocation that would increase the utility of any roommate without decreasing the utility of another.

As another example, the credit assignment problem is solved using an algorithm of de Clippel et al.[14]. Each collaborator reports the relative portion of credit that he believes should be assigned to each of the *other* collaborators. For example, on a project with four collaborators, collaborator A might report that collaborators B and C should receive equal credit while D should receive twice as much credit. The algorithm takes these reports as input and produces a credit assignment that is *impartial*, meaning that an individual's share of credit is independent of his own report, and *consensual*, meaning that if there is a division of credit that agrees with all collaborators' reports then this division is chosen. While these conditions may not sound restrictive, de Clippel et al. [14] show that they are not simultaneously achievable with three collaborators. Their algorithm therefore requires at least four.

In addition to providing a useful set of tools, part of Spliddit's mission is to "communicate to the public the beauty and value of theoretical research in computer science, mathematics, and economics, from an unusual perspective." Indeed, the project has inspired some members of the public to take an interest in algorithms with provable fairness properties. As one example, a representative of one of the largest school districts in California approached the Spliddit team about a problem he was tasked with solving: fairly allocating unused classrooms in public schools to the district's charter schools. This led the Spliddit team, in collaboration with the California school district, to design a practical new approach to classroom allocation that guarantees envy-freeness as well as several other desirable properties [46].

---

[3]http://prediction.microsoft.com/
[4]http://www.spliddit.org/
[5]http://nyti.ms/1o0TUt0

## 3 A Challenge Problem: The Crowdsourcing Compiler

A concrete challenge problem for future research in social computing is what might be called the "Crowdsourcing Compiler":[6] the development of high-level programming languages for specifying large-scale, distributed tasks whose solution requires combining traditional computational and networking resources with volunteer (or paid) human intelligence and contributions. The hypothetical compiler would translate an abstract program into a more detailed organizational plan for machines and people to jointly carry out the desired task. In the same way that today's Java programmer is relieved of low-level, machine-specific decisions (such as which data to keep in fast registers, and which in main memory or disk), the future crowdsourcing programmer would specify the goals of their system, and leave many of the implementation details to the Crowdsourcing Compiler. Such details might include which components of the task are best carried out by machine and which by human volunteers; whether the human volunteers should be incentivized by payment, recognition, or entertainment; how their contributions should be combined to solve the overall task; and so on. While a fully general Crowdsourcing Compiler might well be unattainable, significant progress towards it would imply a much deeper scientific understanding of crowdsourcing than we currently have, which in turn should have great engineering benefits. Noteworthy research efforts which can be viewed as steps on the path to the Crowdsourcing Compiler include Emery Berger's AutoMan Project (http://emeryberger.com/research/automan/) [4], as well as both academic and commercial efforts to automate workflow in crowdsourcing and social computing systems (see e.g., http://groups.csail.mit.edu/uid/turkit/ and http://www.crowdflower.com/).

We note that the organizational schemes in most of the successful crowdsourcing examples to date share much in common. The tasks to be performed (e.g., building an online encyclopedia, labeling images for their content, creating a network of website bookmark labels, finding surveillance balloons) are obviously parallelizable, and furthermore the basic unit of human contribution required is extremely small (fix some punctuation, label an image, etc.). Furthermore, there is usually very little coordination required between the contributions. The presence of these commonalities is a source of optimism for the Crowdsourcing Compiler — so far, there seems to be some shared structure to successful crowdsourcing that the compiler might codify. But are such commonalities present because they somehow delineate fundamental limitations on successful crowdsourcing — or is simply because this is the "low-hanging fruit?"

As of today, the Crowdsourcing Compiler is clearly a "blue sky" proposal meant more to delineate an ambitious research agenda for social computation than it is a guide to short-term steps. But we believe that such an agenda would both need and drive research on theoretical foundations. First steps toward developing the mathematical foundations of a Crowdsourcing Compiler include formally addressing the following questions:

◗ For a given set of assumptions about the volunteer force, and given the nature of the task, what is the best scheme for organizing the volunteers and their contributions? For instance, is it a "flat" scheme where all contributors are equal and their contributions are combined in some kind of majority vote fashion? Or is it more hierarchical, with proven and expert contributors given higher weight and harder subproblems? Which of these (or other) schemes should be used under what assumptions on the nature of the task and what assumptions on the volunteers?

◗ How can we design crowdsourced systems for solving tasks that are much more challenging and less "transactional" than what we currently see in the field — for instance, complex problems where there are strong constraints and interdependencies between the contributions of different volunteers? Behavioral research in recent years has shown that groups of humans can indeed excel on such tasks [44, 45], but we are far from understanding when and why.

---

[6]See http://bit.ly/20juYEX and http://bit.ly/1nIyc3P.

# 4 Challenges to Overcome

We have argued that mathematical research has the potential to make great contributions to social computing. However, before this potential is fully realized, there are several challenges that must be addressed.

## 4.1 Blending Mathematical and Experimental Research

Mathematical and experimental research are complementary and both are needed to develop relevant mathematical foundations for social computing. The strengths of mathematical work include:

1. Mathematical modeling and analysis can be used to cleanly formulate and answer many questions about system behavior without requiring that we build a complete system, providing us with a tool to evaluate the impact of design decisions before committing to any particular design. For example, such models can provide guidance on how to increase participation (e.g., comparing a leaderboard to badges [27, 39]), predict whether a social computing system will achieve critical mass, and perhaps understand how the behavior of groups of users change as the system scales.

2. Mathematical guarantees are desirable for properties like user privacy (which can be obtained, for example, using techniques from the extensive and growing literature on differential privacy [18]), correctness of a system's output, or the scalability of a social computing system.

3. Theoretical work in computer science provides tools for designing and analyzing new algorithms that could lie at the heart of social computing applications, answering questions like how to aggregate noisy and unstructured estimates or information from crowds [36, 43], how to optimally divide a community into subgroups, or how to bring people together in moments of spare time to achieve a common goal.

4. Mathematical models can be used to explore counterfactual analysis, something that is notoriously difficult to do through experiments alone.

Needless to say, mathematical modeling should not and cannot replace experimental work. A mathematical theory can only be truly tested through experiments, and discrepancies between the theory and experimental results provide guidance about how to revise the theory. For example, the ability of mathematical models to make valuable predictions about system behavior depends on an accurate model of system users, which is generally best developed through experimental work.

## 4.2 Learning from the Social Sciences

Computer scientists cannot develop the mathematical foundations of social computing in isolation. Social computing systems are fundamentally *social*. These systems cannot be properly modeled or analyzed without accounting for the behavior of their human components. Much of the literature thus far uses standard models of economic agents and corresponding assumptions about agent preferences, but a growing literature based on experimental work on online platforms suggests that human behavior in several online settings might deviate from these models [38, 49, 67], and these deviations can have significant consequences for how to optimally design social computing systems [20, 27].

In order for mathematical foundations to provide useful practical results, it is necessary to base it on models that better reflect human behavior. This is most effectively achieved via a dialog between theoretical and experimental and empirical research, with studies of human behavior informing mathematical modeling choices, as well as mathematical results suggesting the most important agent characteristics to understand via experimental research. It will be important to understand and incorporate relevant research from psychology, economics, sociology, and other fields. For example, behavioral economics and psychology provide insight into how humans respond to incentives. This is no small task. The best results will be achieved if computer scientists work together with researchers in other fields.

## 4.3 Generalization

Most of the existing mathematical work on social computing focuses on a single application. What does the research on prediction market design tell us about recommendation systems or citizen science? Models will have the most potential for impact if they incorporate reusable components, allowing results to generalize to many systems. (This is one motivation for the Crowdsourcing Compiler of Section 3.)

A related issue is the lack of consensus and understanding of the "core social computing problems," or even if such a set of core problems exists. Mathematical theories are typically developed with one or more such core problems in mind. Such problems should capture challenges that span a wide range of applications and be robust to small changes in the applications to be sure that they are capturing something "real." Clearly, the identification of such problems requires a dialog between practitioners building real systems and theoreticians to identify the most pressing problems requiring mathematical study.

### 4.4 Transparency, Interpretability, and Ethical Implications

One final challenge to overcome is the potential need to make social computing algorithms and models transparent and interpretable to the users of social computing systems. Users are becoming increasingly sophisticated and are aware that the algorithms employed online impact both their day-to-day user experience and their privacy. When faced with the output of an algorithm, many will question where this output came from and why. It is already difficult to explain to users why complex probabilistic algorithms and models produce the results that they do, and this will only become more difficult as algorithms integrate human behavior to a larger extent.

The issue of algorithmic transparency is often tied to ethical concerns such as discrimination and fairness. Examining and avoiding the unintended consequences of opaque decisions made by algorithms is a topic that has been gaining interest in the machine learning and big data communities.[7] Such concerns will undoubtedly need to be addressed in the context of social computing as well.

Acknowledgments. We thank all the workshop participants for their contributions to the workshop, especially to the ideas presented in Section 4. We are grateful to Ashish Goel for his generous help with the content in Section 2, and to Vince Conitzer, David McDonald, David Parkes, and Ariel Procaccia for extensive comments on earlier drafts of this paper.

---

[7]See, for example, the series of recent workshops on Fairness, Accountability, and Transparency in Machine Learning (http://www.fatml.org/).

## Contributors

**Workshop Organizing Committee:**

◗ Yiling Chen, Harvard University

◗ Arpita Ghosh, Cornell University

◗ Tim Roughgarden, Stanford University

◗ Jennifer Wortman Vaughan, Microsoft Research

**CCC Liaisons and Support:**

◗ Lorenzo Alvisi, University of Texas, Austin

◗ Ann Drobnis, Computing Community Consortium

◗ Tal Rabin, IBM Research

◗ Salil Vadhan, Harvard University

◗ Helen Wright, Computing Community Consortium

**Workshop Participants:**

◗ Mitra Basu, National Science Foundation

◗ Shuchi Chawla, University of Wisconsin, Madison

◗ Vince Conitzer, Duke University

◗ Kevin Crowston, Syracuse University

◗ Faisal DSouza, NITRD

◗ Susan R. Fussell, Cornell University

◗ Eric Gilbert, Georgia Tech

◗ Ashish Goel, Stanford University

◗ Bala Kalyanasundaram, Georgetown University

◗ Michael Kearns, University of Pennsylvania

◗ Tracy Kimbrel, National Science Foundation

◗ Joe Konstan, University of Minnesota

◗ Ee-Peng Lim, Singapore Management University

◗ Keith Marzullo, OSTP/NCO

◗ Winter Mason, Facebook

◗ David McDonald, University of Washington

◗ Nina Mishra, Amazon

◗ Elizabeth Mynatt, Georgia Tech and CCC

◗ Tristan Nguyen, Air Force Office of Scientific Research

◗ Lynne Parker, National Science Foundation

◗ David Parkes, Harvard University

◗ Sharoda Paul, GE Global Research

◗ Sid Suri, Microsoft Research

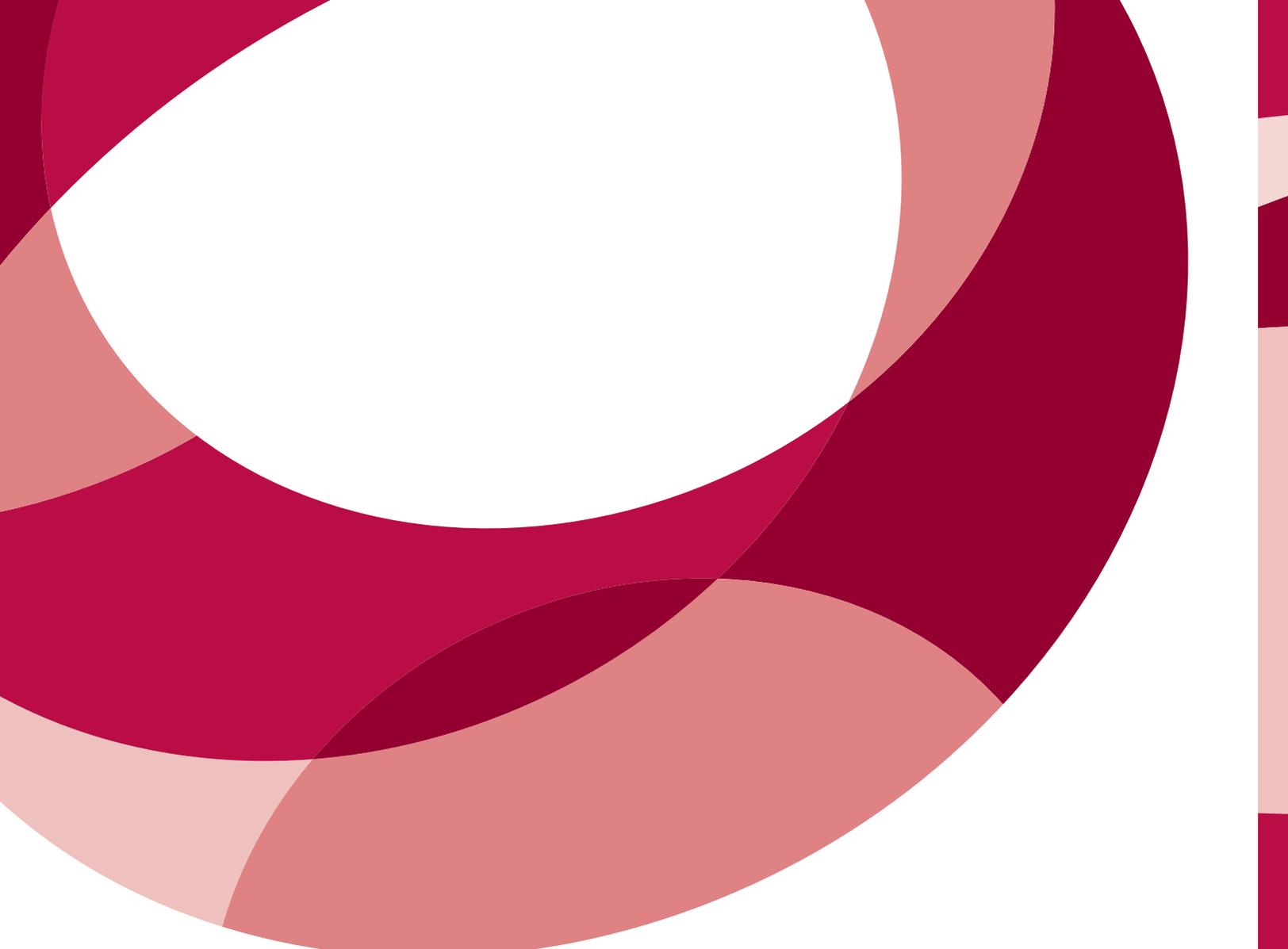

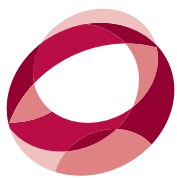

CCC
Computing Community Consortium
Catalyst

1828 L Street, NW, Suite 800
Washington, DC 20036
P: 202 234 2111 F: 202 667 1066
www.cra.org cccinfo@cra.org